\documentclass[preprint,prd,aps,showpacs,showkeys,nofootinbib]{revtex4}

\def\XXint#1#2#3{{\setbox0=\hbox{$#1{#2#3}{\int}$}
     \vcenter{\hbox{$#2#3$}}\kern-.5\wd0}}

\usepackage{graphicx}
\usepackage{bm}
\usepackage{amssymb}
\usepackage{amsmath}
\usepackage{euscript}

\begin{document}

\title{The fundamental $1/f$ noise in monolayer graphene}

\author{Kirill~A.~Kazakov}

\affiliation{Department of Theoretical Physics,
Physics Faculty,\\
Moscow State University, $119991$, Moscow, Russian Federation}

\begin{abstract}
The quantum indeterminacy caused by non-commutativity of observables at different times sets a lower bound on the voltage noise power spectrum in any conducting material. This bound is calculated explicitly in the case of monolayer graphene. It is found that account of graphene pseudospin/valley band structure raises the quantum bound by a factor of $7/2$ compared to the case of spinless charge carriers with a conical energy-momentum dispersion. The bound possesses all characteristic properties of $1/f$ noise, and its dependence on the charge carrier density is congruent to the experimentally observed.
\end{abstract}

\maketitle

Among numerous fluctuation phenomena found in conducting media, the $1/f$ (flicker) noise is by far the least understood despite its ubiquity and practical importance. This notion covers all electrical fluctuations whose power spectral density $S(f)\sim 1/f^{\gamma}$ at low frequencies $f,$ where the frequency exponent $\gamma$ is around unity \cite{johnson,buckingham,bell1980,raychaudhuri}. On the one hand, this noise limits performance of microelectronic devices, but on the other hand, measurements of the $1/f$ spectrum are widely used to identify certain physical processes in various materials and to assess structural properties of samples. An adequate characterization of the possible $1/f$-spectra is thus vital for these applications.

It is generally accepted that $1/f$ noise is due to fluctuations in electrical conductivity caused by various physical processes such as charge carrier trapping, temperature fluctuations, generation-recombination of carriers in semiconductors, impurity migration, {\it etc.} An important common feature of the corresponding power spectra is that they all eventually flatten at sufficiently low frequencies, so that their $1/f$-components cover only a few frequency decades. At the same time, $1/f$ noise has been experimentally detected down to frequencies as low as $f=10^{-6.3}$ Hz, with no sign of a low-frequency cutoff \cite{rollin1953,caloyannides}. As this is far below of any characteristic frequency of the aforementioned conventional physical processes, one is led to the problem of origin of {\it fundamental flicker noise}, which is the component of $1/f$ spectrum presumably extending down to $f=0.$ It was suggested long ago by P.~Handel that its origin might be in the well-known quantum infrared divergence in the coupling of electric charges to photons \cite{handel1,handel2}. Irrespective of the severe critics that followed Handel's derivation, his theory predicts that the frequency exponent is strictly less than unity, $1 > \gamma = 1 - O(\alpha),$ where $\alpha \approx 1/137$ is the fine structure constant. Though the deviation from unity is tiny, it ensures convergence of the total noise power in Handel's theory \cite{handel1,handel7}. However, the numerous flicker noise measurements unequivocally demonstrate that $(\gamma-1)$ can be positive as well as negative (it is actually predominantly positive, as {\it e.g.}, in Ref.~\cite{caloyannides}).

It is the case $\gamma>1$ that makes the problem of fundamental noise so acute, because of an apparent conflict between the observed absence of a low-frequency cutoff and finiteness of the noise variance. In fact, a direct consequence of the celebrated Wiener-Khinchin relation \cite{wiener,khinchin} is that this variance is equal to $2\int_{0}^{\infty} {\rm d}f S(f),$ which is formally infinite because the integral diverges at $f=0$ whenever $S(f)\sim 1/f^{\gamma}$ with $\gamma>1.$ Yet, it was shown recently that account of the quantum nature of charge carrier interaction with photons does resolve this paradox, though in a way quite different from Handel's \cite{kazakov1,kazakov2}. First of all, it turns out that quantum theory implies certain limitations on applicability of the Wiener-Khinchin relation to the voltage measurements, and second, it sets a lower bound on the spectral power of voltage fluctuations which possesses all characteristic features of the observed $1/f$ noise. In Refs.~\cite{kazakov1,kazakov2} this bound was calculated explicitly for materials with a parabolic energy-momentum dispersion of charge carriers, which allowed a direct comparison of the result with experiment. Such a comparison with $1/f$ noise measured in InGaAs quantum wells and high-temperature superconductors has revealed that the observed noise levels are only a few times as high as the bound established.

An important property of the quantum bound is its inverse proportionality to the charge carrier mass. This raises a curious question of what the noise magnitude is in the case of conical energy-momentum dispersion (massless charge carriers). The answer is that the zero-mass singularity of the power spectrum for parabolic dispersion transmutes into a zero-momentum singularity of its decomposition with respect to the charge carrier momentum \cite{kazakov3}. A measurable consequence of this singularity is a sharp peak in the noise magnitude at small charge carrier density. This result was applied in Ref.~\cite{kazakov3} to monolayer graphene, in which case the peak becomes M-shaped on account of a continuous transition from the electron to hole conductivity. A comparison with experimental data in GraFETs has shown that the calculated power spectrum, considered as a function of the charge-carrier density, is congruent to the observed, though the excess of the measured noise magnitudes over the bound is more pronounced than in the case of parabolic dispersion. On the other hand, this comparison was only qualitative, as accurate as a mere replacement of the dispersion law permits. To obtain a quantitatively reliable prediction, and in particular, to decide if graphene really exhibits excessive noise, calls for a more careful account of the underlying physics. The purpose of this Letter is to make a first step in this direction and to obtain the quantum bound taking into account the charge carrier pseudospin.

Before going into details of the calculation, it is useful to summarize the main points of the new approach. Consider a (semi)conducting sample biased so that a constant-on-average electric field $\bm{E}$ is established within, and let the voltage across the sample be measured by means of two probes assumed for simplicity pointlike, $\bm{x}_1,\bm{x}_2$ denoting their position. A voltage measured at time $t$ is the sum of a mean value $U_0(\bm{x}_1,\bm{x}_2)$ and a fluctuation, or noise, $\Delta U(t,\bm{x}_1,\bm{x}_2).$ The voltage fluctuation $\Delta U(t,\bm{x}_1,\bm{x}_2)$ is an observable to which there corresponds a Hermitian (Heisenberg) operator $\widehat{\Delta U}(t)$ (for brevity, the arguments $\bm{x}_1,\bm{x}_2$ in $\widehat{\Delta U}$ are henceforth omitted). According to Refs.~\cite{kazakov1,kazakov2}, the quantum bound on the power spectral density of voltage fluctuations, $S_F(f),$ arises from the quantum indeterminacy caused by non-commutativity of observables at different times. It is this indeterminacy that obstructs applicability of the Wiener-Khinchin relation to voltage fluctuations, which in turn is the reason why the quantum bound needs no low-frequency cutoff. In order to find $S_F(f),$ one first defines a correlation function $S(t-t') = \left\langle\widehat{\Delta U}(t)\widehat{\Delta U}(t')\right\rangle,$ where angular brackets denote the quantum-mechanical average over a fixed state of the system ``conducting sample plus electromagnetic field.'' $S(t-t')$ can be written as the sum of two terms, one of which is symmetric under the interchange $t\leftrightarrow t',$ and the other is antisymmetric. As was demonstrated in Refs.~\cite{kazakov1,kazakov2}, the latter admits a Fourier decomposition despite the presence of $1/f$-term in the power spectrum, and defines $S_F(f)$ according to
\begin{eqnarray}\label{sigma}
S_F(f) = \left|\lim\limits_{t_m\to \infty}\left\{\int_{-t_m}^{t_m}{\rm d}\tau S(\tau)\sin(\omega\tau) - \frac{1}{t_m}\int_{-t_m}^{t_m}{\rm d}\tau |\tau|S(\tau) \sin(\omega\tau)\right\}\right|, \quad \omega = 2\pi f.
\end{eqnarray} It is a key fact of the present approach that the limit under the modulus sign in Eq.~(\ref{sigma}) exists even when $S(\tau)\sim \tau^{\gamma-1}$ for $\tau \to \infty,$ as is the case when $S(f)\sim 1/f^{\gamma}$ at low frequencies.

A straightforward way to compute the expectation value $\left\langle\widehat{\Delta U}(t)\widehat{\Delta U}(t')\right\rangle$ is to use the Schwinger-Keldysh technique \cite{schwinger,keldysh}, according to which it
is written as
\begin{eqnarray}\label{expectation2}
\left\langle\widehat{\Delta U}(t)\widehat{\Delta U}(t')\right\rangle = {\rm tr}\left(\hat{\rho}\,\EuScript{T}_C\widehat{\Delta u}^{(2)}(t)\widehat{\Delta u}^{(1)}(t')\exp\left\{-{\rm i}\int_{C}{\rm d}t\,\hat{w}(t)\right\}\right),
\end{eqnarray}
where $\widehat{\Delta u}(t) = \hat{a}_{0}(t,\bm{x}_1)-\hat{a}_{0}(t,\bm{x}_2) - U_0,$ with $\hat{a}_{0}$ the scalar electromagnetic potential, $\hat{\rho}$ is the system density matrix, $\hat{w}$ is the interaction Hamiltonian, and the lowercase letters denote operators in the interaction picture; the so-called Schwinger-Keldysh time contour $C$ runs from $t=-\infty$ to $t=+\infty$, and then back to $t=-\infty,$ with all time instants on the forward branch [designated with a superscript $(1)$] treated as being in the past with respect to any time instant on the backward branch [designated with a superscript $(2)$]; $\EuScript{T}_C$ denotes operator ordering along this contour.

Let us now specify physical conditions under which the power spectrum of voltage fluctuations in graphene will be evaluated. As is well known, the conical energy-momentum dispersion of charge carriers is found in monolayer graphene \cite{geim2009}. Therefore, the sample is a planar monolayer of carbon atoms arranged in a hexagonal lattice. The energy-momentum dispersion is approximately conical near the so-called Dirac points which are two corners of the graphene Brillouin zone \cite{wallace1947}. In the approximation where the electron hopping between nearest sites of the two triangular sublattices is only taken into account, the effective free electron Hamiltonian reads \cite{semenoff1984}
\begin{eqnarray}\label{hamiltonian}
\hat{H}_0 = v_F\int {\rm d}^2\bm{x}\left[\hat{\Psi}_1^{\dagger}(\bm{x})\bm{\sigma}\cdot(-{\rm i}\hbar\bm{\nabla})\hat{\Psi}_1(\bm{x}) + \hat{\Psi}_2^{\dagger}(\bm{x})\bm{\sigma}^*\cdot(-{\rm i}\hbar\bm{\nabla})\hat{\Psi}_2(\bm{x})\right],
\end{eqnarray} where the matrix vectors $\bm{\sigma} = (\sigma_x,\sigma_y),$ $\bm{\sigma}^* = (\sigma_x,-\sigma_y),$ $\sigma_x,\sigma_y$ being the standard Pauli matrices in the sample plane, and $\hat{\Psi}_{1,2}(\bm{x})$ are the (pseudo)spinor electron fields, the subscripts 1,2 referring to the two Dirac points. Thus, the electrons are massless Dirac-like particles with the energy-momentum dispersion $\varepsilon(\bm{q})=\pm v_F|\bm{q}|$ (Dirac point being at $\bm{q}=0$), so that the charge carriers include two copies of electrons ($\varepsilon>0$) and holes ($\varepsilon<0$), all characterized by the same value of Fermi velocity $v_F\approx 10^8\,$cm/s. Except in the number of states counting, the ordinary spin of charge carriers does not play a role in the present consideration, and so the charge carriers are assumed unpolarized, with all spin indices accordingly suppressed. At last, also for simplicity, all finite-temperature effects are neglected. From the experimental standpoint, dependence of the power spectrum on the charge carrier density is of major interest. As a means to control this density, it is best to consider the graphene sample as part of a field-effect transistor -- GraFET.

Interaction with the external fields is introduced, as usual, by invoking the principle of gauge invariance to replace the ordinary space-time derivatives by covariant ones in the Schr$\ddot{\rm o}$dinger equation. This yields the interaction Hamiltonian, in the interaction picture,
\begin{eqnarray}\label{inthamiltonian}
\hat{w}(t) = \int {\rm d}^2\bm{x}\left\{v_F\frac{e}{c}\left[\hat{\psi}_1^{\dagger}\bm{\sigma}
\cdot\hat{\bm{a}}\hat{\psi}_1 + \hat{\psi}_2^{\dagger}\bm{\sigma}^*\cdot\hat{\bm{a}}\hat{\psi}_2\right]+ e\hat{a}_0[\hat{\psi}_1^{\dagger}\hat{\psi}_1 + \hat{\psi}_2^{\dagger}\hat{\psi}_2]\right\},
\end{eqnarray}
where $\hat{\bm{a}}$ is the electromagnetic vector potential. Since the charge carriers are non-relativistic ($v_F \ll c,$ the speed of light in vacuum), $\hat{\bm{a}}$ in Eq.~(\ref{inthamiltonian}) can be replaced by the classical vector potential of the electric field, $\bm{E},$ established in the sample,
\begin{eqnarray}\label{avector}
{\bm a}(t,\bm{x}) = {\rm i}\bm{E}(\bm{x})\frac{{\rm e}^{{\rm i\lambda t}} - 1}{\lambda}\,, \quad \lambda \to 0.
\end{eqnarray}

\begin{figure}
\includegraphics[width=0.75\textwidth]{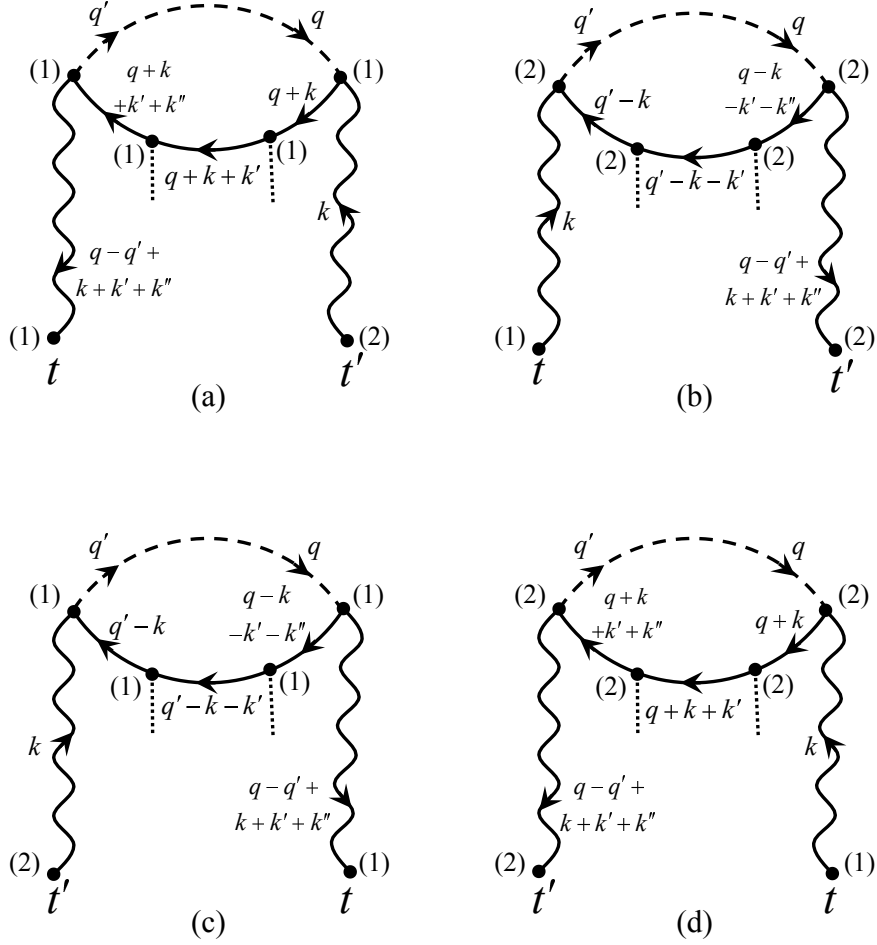}
\caption{Non-vanishing diagrams representing $\left\langle\widehat{\Delta U}(t)\widehat{\Delta U}(t')\right\rangle$ in the Schwinger-Keldysh formalism. Solid and wavy lines depict propagators of the charge carriers and  electromagnetic field, respectively, dashed line -- charge carrier density matrix, dotted line -- external electric field. The arrows on the lines show the energy-momentum flow.}\label{fig1}
\end{figure}

The set of lowest-order nonvanishing Schwinger-Keldysh diagrams contributing to the $1/f$ asymptotic of the power spectrum are drawn in Fig.~\ref{fig1}. Each diagram involves two vertices of the charge carrier interaction with the external electric field. As was found in Ref.~\cite{kazakov3}, similar diagrams generated by the model Hamiltonian that neglects the electron pseudospin cancel each other, so that the non-trivial contribution comes only from diagrams involving vertices quadratic in $\bm{E}.$ In contrast, there are no such vertices in the interaction Hamiltonian (\ref{inthamiltonian}), but the sum of diagrams in Fig.~\ref{fig1} no longer vanishes because of the charge carrier pseudospin fully taken into account by this Hamiltonian. In any case, the power spectrum in the lowest, forth order with respect to the electron charge is quadratic with respect to $\bm{E}.$ On the other hand, a consistent evaluation of the higher order corrections would require explicit account of the charge carrier collisions \cite{kazakov2}.

The sum of diagrams in Fig.~\ref{fig1} reads, in relativistic units $\hbar=c=1$ (it is doubled to account for the equal contributions of the fields $\psi_1$ and $\psi_2$)
\begin{eqnarray}\label{longhand}
S(\tau) &=& v^2_F(4\pi)^2 e^4\frac{\partial^2}{\partial\lambda'
\partial\lambda''}\int\frac{{\rm d}^4 k}{(2\pi)^4}\frac{{\rm d}^3 \bm{q}}{(2\pi)^3}\frac{{\rm d}^3 \bm{q}'}{(2\pi)^3}\varrho(\bm{q},\bm{q}')\left({\rm e}^{{\rm i}\bm{k}\cdot(\bm{x}_1 - \bm{x}_2)} - 1\right) {\rm e}^{{\rm i}(\bm{q}-\bm{q}')\cdot\bm{x}_1 }\nonumber\\&& \times\left\{{\rm e}^{-{\rm i}k^0\tau-{\rm i}(\lambda'+\lambda'')t}G^{(11)}(q-q'+k+k'+k'')G^{(12)}(k)
\right.\nonumber\\&&\left.\hskip0,6cm\times{\rm tr}\left[D^{(11)}(q+k+k'+k'')\bm{\sigma}\cdot\bm{E}D^{(11)}(q+k+k')
\bm{\sigma}\cdot\bm{E}D^{(11)}(q+k) \right.\right.\nonumber\\&& \left.\left.\hskip0,6cm + D^{(11)}(q'-k)\bm{\sigma}\cdot\bm{E}
D^{(11)}(q'-k-k')\bm{\sigma}\cdot\bm{E}D^{(11)}(q'-k-k'-k'')\right]
\right.\nonumber\\&& \left.\hskip0,6cm + {\rm e}^{{\rm i}k^0\tau-{\rm i}(\lambda'+\lambda'')t'}G^{(22)}(q-q'+k+k'+k'')G^{(21)}(k)
\right.\nonumber\\&& \left.\hskip0,6cm\times {\rm tr}\left[ D^{(22)}(q+k+k'+k'')\bm{\sigma}\cdot\bm{E}D^{(22)}(q+k+k')
\bm{\sigma}\cdot\bm{E}D^{(22)}(q+k)
\right.\right.\nonumber\\&& \left.\left.\hskip0,6cm + D^{(22)}(q'-k)\bm{\sigma}\cdot\bm{E}D^{(22)}(q'-k-k')
\bm{\sigma}\cdot\bm{E}D^{(22)}(q'-k-k'-k'')\right]
\right\} \Bigr|_{\lambda=\lambda'=0} \nonumber\\ && + (\bm{x}_1 \leftrightarrow\bm{x}_2),
\end{eqnarray} where a 4-vector notation is used, $k=(k^0,\bm{k}), q=(\varepsilon(\bm{q}),\bm{q}),$ $k' = (\lambda',\bm{0}),$ $k'' = (\lambda'',\bm{0});$ $G$ and $D$ are the momentum-space Schwinger-Keldysh propagators of temporal photons and charge carriers, respectively,
\begin{eqnarray}\label{photonprop}
G^{(11)}(k) &=& \frac{{\rm i}}{k^2 + {\rm i}0} = [G^{(22)}(k)]^*, \quad G^{(12)}(k) = 2\pi\theta(-k^0)\delta(k^2), \quad G^{(21)}(k) = 2\pi \theta(k^0)\delta(k^2),\nonumber \\
D^{(11)}(p) &=& {\rm i}\left\{p^0 - v_F\bm{\sigma}\cdot\bm{p}+{\rm i}0\right\}^{-1} = {\rm i}\frac{p^0 + v_F\bm{\sigma}\cdot\bm{p}}{(p^0+{\rm i}0)^2 - v^2_F\bm{p}^2}\,, \quad  D^{(22)}(q) = -{\rm i}\frac{p^0 + v_F\bm{\sigma}\cdot\bm{p}}{(p^0 - {\rm i}0)^2 - v^2_F\bm{p}^2 }\,,\nonumber
\end{eqnarray} and the step function $\theta(k^0)=0$ for $k^0 \leqslant 0$, $\theta(k^0)=0$ for $k^0>0$; the trace is over the charge carrier pseudospin; since the photon heat bath effects are neglected, $G$'s are purely vacuum; accordingly, the system density matrix $\hat{\rho}$ is reduced to that of the charge carriers, $\varrho(\bm{q},\bm{q}')$ denoting its momentum-space representation; the latter is normalized by
\begin{eqnarray}\label{normalization1}
\int\frac{{\rm d}^3 \bm{q}}{(2\pi)^3}\varrho(\bm{q},\bm{q}) = 1.
\end{eqnarray}
At last, ``$+ (\bm{x}_1 \leftrightarrow\bm{x}_2)$'' means that the preceding expression is to be added with $\bm{x}_1$ and $\bm{x}_2$ interchanged.

To extract the low-frequency asymptotic of the power spectrum, we note that at frequencies $k^0$ typical for the flicker noise studies, the photon 4-momentum $k$ associated with the propagators $G^{(12)},G^{(21)}\sim \delta(k^2)$ is negligibly small compared to that of the charge carriers. With a great accuracy, therefore, denominators of the charge carrier propagators can be simplified as
$$(q^0+k^0)^2 - v^2_F(\bm{q}+\bm{k})^2 \approx 2v_F|\bm{q}|k^0,$$ {\it etc.} Next, the traced expressions in the numerator are of the form
\begin{eqnarray}&&
{\rm tr}\left[(s^0 + v_F\bm{\sigma}\cdot\bm{s})\bm{\sigma}\cdot\bm{E}(s^0 \pm \lambda' + v_F\bm{\sigma}\cdot\bm{s})\bm{\sigma}\cdot\bm{E}(s^0 \pm \lambda' \pm \lambda'' + v_F\bm{\sigma}\cdot\bm{s})\right] \nonumber\\&&
= 2\bm{E}^2s^0(s^0\pm\lambda')(s^0 \pm \lambda' \pm \lambda'')- 2v^2_F\bm{E}^2\bm{s}^2(s^0 \pm \lambda') + 4v^2_F(\bm{s}\cdot\bm{E})^2(2s^0 \pm \lambda' \pm \lambda''),\nonumber
\end{eqnarray} where $s=q+k, q'-k.$ Neglecting, as before, $\bm{k}$ in comparison with $\bm{q},\bm{q}'$ and assuming for simplicity that the momentum distribution of charge carriers is isotropic in the sample plane, $(\bm{s}\cdot\bm{E})^2\approx (\bm{q}\cdot\bm{E})^2$ in the last term can be replaced by $\bm{q}^2\bm{E}^2/2$ (though the factor ${\rm e}^{{\rm i}(\bm{q}-\bm{q}')\cdot\bm{x}_1 }$ in the integrand of Eq.~(\ref{longhand}) explicitly breaks the isotropy, it is not difficult to see from the subsequent calculations that this only adds terms which vanish unless the charge carrier is localized on a voltage probe). Written longhand, expression (\ref{longhand}) thus takes the form
\begin{eqnarray}\label{longhand1}
S(\tau) &=& \frac{4\pi^2e^4\bm{E}^2}{v_F} \frac{\partial^2}{\partial\lambda'
\partial\lambda''}\int\frac{{\rm d}^4 k}{(2\pi)^3}\frac{{\rm d}^3 \bm{q}}{(2\pi)^3}\frac{{\rm d}^3 \bm{q}'}{(2\pi)^3}\varrho(\bm{q},\bm{q}')\left({\rm e}^{{\rm i}\bm{k}\cdot(\bm{x}_1 - \bm{x}_2)} - 1\right)
\nonumber\\&& \times {\rm e}^{{\rm i}(\bm{q}-\bm{q}')\cdot\bm{x}_1 }\delta(k^2)\frac{{\rm e}^{-{\rm i}k^0\tau-{\rm i}(\lambda'+\lambda'')t}\theta(-k^0)+{\rm e}^{{\rm i}k^0\tau-{\rm i}(\lambda'+\lambda'')t'}\theta(k^0)}
{(q-q')^2k^0(k^0+\lambda')(k^0+\lambda'+\lambda'')}
\nonumber\\&& \times
\left\{\left(v_F+\frac{k^0+\lambda'}{|\bm{q}|}\right)
\left[\left(v_F+\frac{k^0}{|\bm{q}|}\right)
\left(v_F+\frac{k^0+\lambda'+\lambda''}{|\bm{q}|}\right)+ v^2_F\right]
 \right.\nonumber\\&& \left. \hskip0,5cm -\left(v_F-\frac{k^0+\lambda'}{|\bm{q}'|}\right)
\left[\left(v_F-\frac{k^0}{|\bm{q}'|}\right)
\left(v_F-\frac{k^0+\lambda'+\lambda''}{|\bm{q}'|}\right)+ v^2_F\right]\right\}\Biggl|_{\lambda=\lambda'=0} \nonumber\\ && + (\bm{x}_1 \leftrightarrow\bm{x}_2).
\end{eqnarray}
Since $|\bm{k}|=|k^0|$ (because of the factor $\delta(k^2)$), the integrand is to be expanded with respect to $\bm{k}$ as well as to $k^0$ keeping only the leading term, {\it e.g.}, $\exp[{\rm i}\bm{k}\cdot(\bm{x}_1 - \bm{x}_2)] - 1 = - [\bm{k}\cdot(\bm{x}_1 - \bm{x}_2)]^2/2.$ Taking into account also that the expression in the braces is $O(k^0)$ at $\lambda'=\lambda''=0,$ a simple power counting shows that the integrand is $O(1/k^0)$ for $k^0 \to 0,$ as expected. It is also readily checked that $\lambda'$- or $\lambda''$-differentiation of the time-dependent exponential factors gives rise to terms which are non-singular at $k^0=0.$ These contribute only to the symmetric part of $S(\tau)$ and are ill-defined, as this part lacks temporal Fourier decomposition \cite{kazakov2}. On the other hand, the $k^0$-integral in the antisymmetric part is well-defined. A straightforward calculation yields

\begin{eqnarray}\label{lowest}
S(\tau) =&& \frac{14v_Fe^4\bm{E}^2(\bm{x}_1 - \bm{x}_2)^2}{3}\int_{0}^{\infty}{\rm d}k^0\frac{{\rm e}^{{\rm i}k^0\tau}}{k^0}\int\frac{{\rm d}^3 \bm{q}}{(2\pi)^3}\frac{{\rm d}^3 \bm{q}'}{(2\pi)^3}\nonumber\\&&\times\varrho(\bm{q},\bm{q}')
\left(\frac{1}{|\bm{q}|} + \frac{1}{|\bm{q}'|}\right)\frac{{\rm e}^{{\rm i}(\bm{q}-\bm{q}')\cdot\bm{x}_1 } + {\rm e}^{{\rm i}(\bm{q}-\bm{q}')\cdot\bm{x}_2 }}{(\bm{q} - \bm{q}')^2}\,.
\end{eqnarray} Following Refs.~\cite{kazakov1,kazakov2}, this expression can be made more specific by expressing the charge carrier density matrix via the mixed position-momentum distribution function $R(\bm{r},\bm{Q}),$
\begin{eqnarray}\label{mixed}
\varrho\left(\bm{Q}-\frac{\bm{p}}{2},\bm{Q}+\frac{\bm{p}}{2}\right) = \frac{1}{\Omega}\int_{\Omega} {\rm d}^3\bm{r}{\rm e}^{{\rm i}\bm{p}\cdot\bm{r}}R(\bm{r},\bm{Q}),
\end{eqnarray} where $\Omega$ is the sample volume. $R(\bm{r},\bm{Q})$ vanishes for $\bm{r}$ outside of the sample, hence, $(\bm{q}' - \bm{q}) = \bm{p}$ is of the order of the inverse linear sample size which in practice is much larger than the lattice constant. Therefore, $|\bm{p}| \ll |\bm{q}|$ for all relevant charge carrier momenta, and so one can set $(1/|\bm{q}| + 1|\bm{q}'|) \approx 2/|\bm{Q}|.$ The integral over $\bm{p}$ in Eq.~(\ref{lowest}) is then just a Fourier decomposition of the Coulomb potential. Finally, the power spectrum of voltage fluctuations is found by substituting an odd in $\tau$ part of $S(\tau)$ into Eq.~(\ref{sigma}). Replacing $\bm{E}^2(\bm{x}_1 - \bm{x}_2)^2$ by $U^2_0$ and restoring ordinary units gives
\begin{eqnarray}\label{sf1}
S_F(f) =&& \frac{7v_Fe^4U^2_0}{6\pi \hbar c^3 |f|}\frac{1}{\Omega}\int_{\Omega} {\rm d}^3\bm{r}\left(\frac{1}{|\bm{r}-\bm{x}_1|} + \frac{1}{|\bm{r}-\bm{x}_2|}\right)\int\frac{{\rm d}^3 \bm{Q}}{(2\pi)^3}\frac{R(\bm{r},\bm{Q})}{|\bm{Q}|}\,.
\end{eqnarray} In the practically important case of macroscopically homogeneous sample, $R(\bm{r},\bm{Q}) = R(\bm{Q}),$ this expression simplifies to:
\begin{eqnarray}\label{sf3}
S_F(f) = \frac{\varkappa U^2_0}{|f|}\,, \quad \varkappa =  \frac{7v_Fe^4 g}{2\pi\hbar c^3}\int\frac{{\rm d}^2 \bm{Q}}{(2\pi\hbar)^2} \frac{R(\bm{Q})}{|\bm{Q}|}\,,
\end{eqnarray}
where the geometric factor $g$ is defined following Ref.~\cite{kazakov1},
$$g = \frac{1}{3\Omega}\int_{\Omega} {\rm d}^3\bm{r}\left(\frac{1}{|\bm{r}-\bm{x}_1|} + \frac{1}{|\bm{r}-\bm{x}_2|}\right).$$ Except for the factor $7/2,$ Eq.~(\ref{sf3}) coincides with the expression derived in Ref.~\cite{kazakov3}. Thus, account of the detailed band structure of graphene raises the lower bound on the power spectrum of flicker noise by a factor of $7/2=2\times (7/4),$ where the factor $2$ is merely due to doubling of charge carrier types, and the rest is specific to the charge carrier pseudospin.

\begin{figure}
\includegraphics[width=0.48\textwidth]{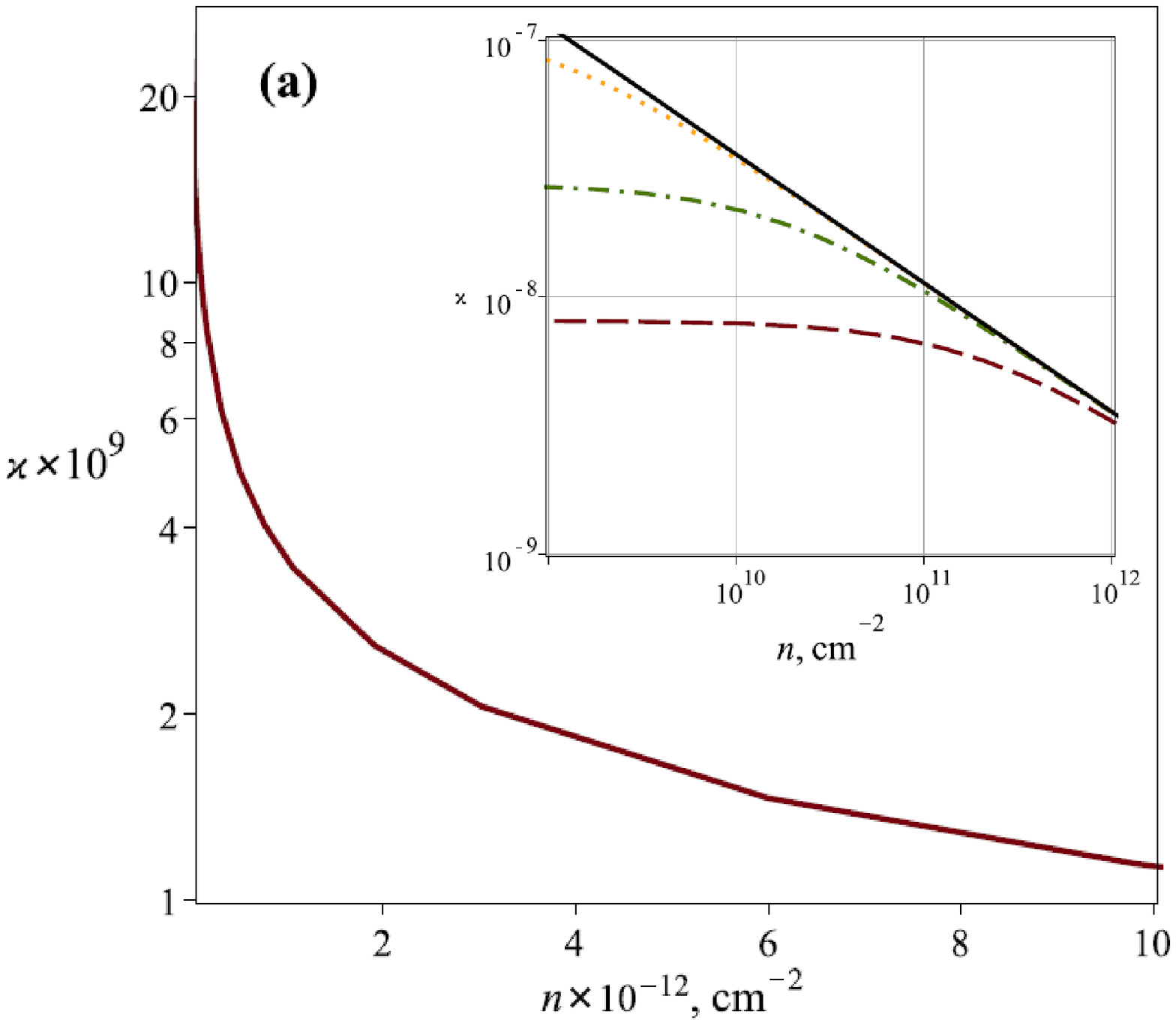}
\includegraphics[width=0.48\textwidth]{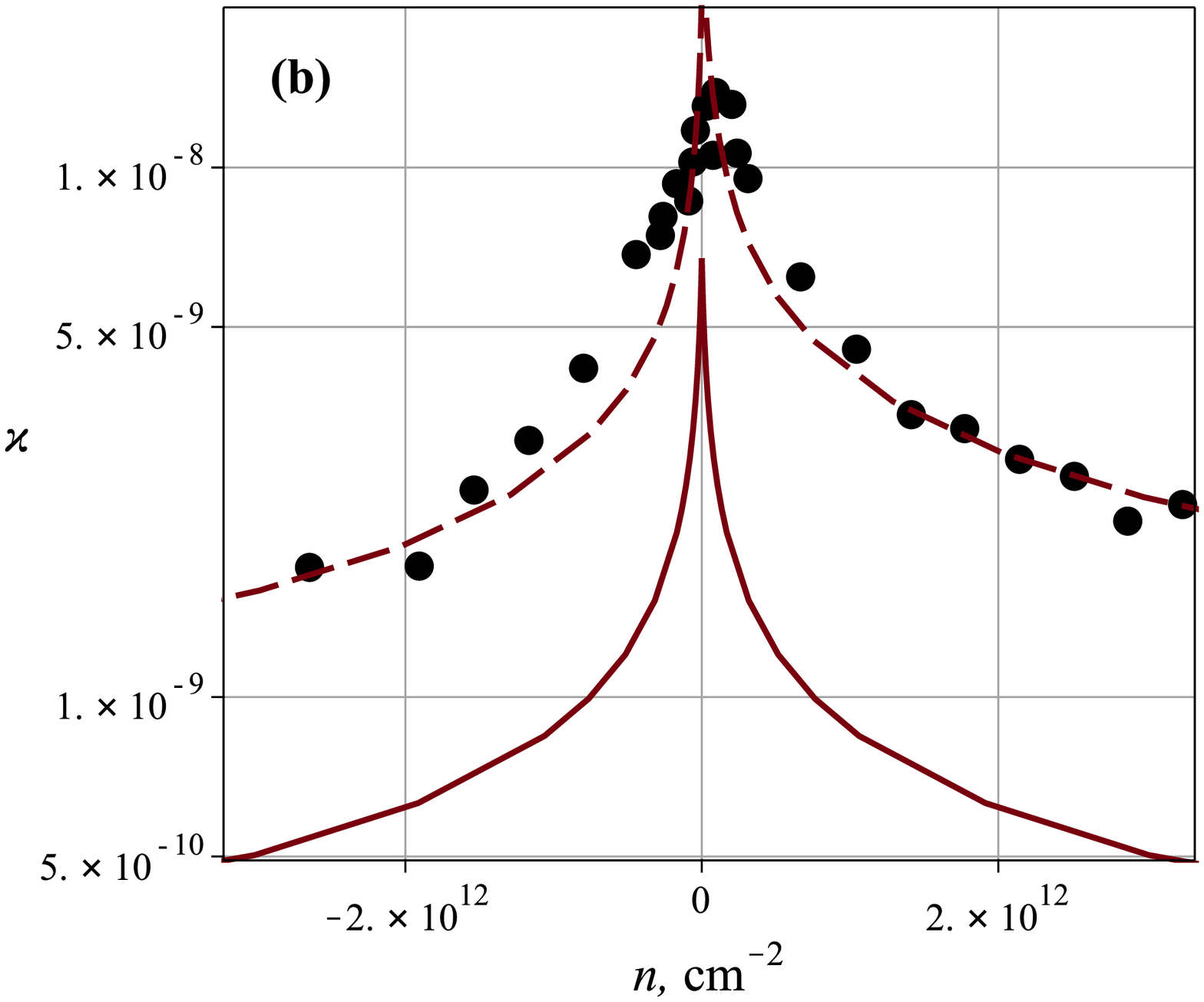}
\caption{(a) A semi-logarithmic plot of the reduced power spectrum, $\varkappa,$ versus charge carrier density, $n,$ suggesting a singularity of $\varkappa$ at $n=0$  (sample size $1\times 1\,$$\mu$m$^2,$ $g\approx 1.6\times 10^4\,$cm$^{-1}$). {\it Inset}: logarithmic plots of $\varkappa(n)$ at various temperatures showing that $\varkappa(0)$ is actually finite. $T=300\,$K (dash), $T=88\,$K (dash-dots), and $T=20\,$K (dots); solid line is the expression (\ref{sf4}). (b) $\varkappa(n)$ measured in Ref.~\cite{pal2011} in a field-effect device SLG1 (marks, as read off from figure 1(d) of Ref.~\cite{pal2011}), and calculated according to Eq.~(\ref{sf4}) (solid line). $n>0$ ($n<0$) is the electron (hole) doping. Dashed curve is obtained by a vertical shift of the solid curve to ease comparison with the experimental data (in view of the observed electron-hole asymmetry, the two wings of the curve were shifted independently).}\label{fig2}
\end{figure}

For the purposes of comparison with experiment, the reduced power spectrum, $\varkappa,$ is to be expressed as a function of the charge carrier density, $n.$ Its analytical expression is easily derived under the assumption that thermal effects are negligible. According to Eqs.~(\ref{normalization1}), (\ref{mixed}), $R(\bm{Q})$ is normalized by
\begin{eqnarray}\label{normalization}
\int\frac{{\rm d}^2 \bm{Q}}{(2\pi \hbar)^2}R(\bm{Q}) = 1.
\end{eqnarray} At zero temperature, one has $R(\bm{Q})=4\theta(q_F-|\bm{Q})|)/n,$ where $q_F$ is the Fermi momentum (the factor of four accounts for the electron spin and the two Dirac points). Substitution of this relation in Eqs.~(\ref{sf3}), (\ref{normalization}) yields
$$\varkappa = \frac{7v_Fe^4 gq_F}{\pi^2\hbar^3 c^3n}\,, \quad n = \frac{q^2_F}{\pi\hbar^2},$$ hence,
\begin{eqnarray}\label{sf4}
\varkappa = \frac{7v_Fe^4 g}{\pi^{3/2}\hbar^2 c^3}n^{-1/2}\,.
\end{eqnarray}
The unbounded growth of $\varkappa$ at small $n$ is a reflection of the singularity at $\bm{Q}=0$ of the integrand in Eq.~(\ref{sf3}).  Account of the thermal effect on the charge carrier distribution removes this singularity, in which case Eq.~(\ref{sf4}) plays the role of a large-$n$ asymptotic of this function. However, the value $n=0$ is unattainable in practice anyway because of the fluctuations in $n$ caused by disorder and charged impurities in the substrate. The corresponding characteristic density fluctuation, $\delta n,$ depends essentially on the charged impurity concentration, and ranges normally $10^{11}\,$cm$^{-2}$ to $10^{12}\,$cm$^{-2}$ \cite{hwang2007}. To give an idea of how Eq.~(\ref{sf4}) applies at various temperatures, it is compared in Fig.~\ref{fig2}(a) with the function $\varkappa(n)$ evaluated taking into account thermal effect on the charge carrier distribution \cite{kazakov3}.

The correction factor $7/2$ found above is important for identifying the origin of flicker noise in graphene, as it shifts the quantum bound closer to the experimentally observed values. This is illustrated in Fig.~\ref{fig2}(b) where the function $\varkappa(n)$ calculated according to Eq.~(\ref{sf4}) is compared to the measured in a sample of size $3.5\times 4.5\,$$\mu$m$^2$ \cite{pal2011} (in which case $g\approx 4\times 10^3\,$cm$^{-1}$). It is seen that the two sets of values are nearly congruent, though the experimental values are three to five times as large as the quantum bound. One might think that this discrepancy in the magnitude rules out the quantum indeterminacy as a dominant noise source. However, there is actually another important factor affecting the noise magnitude which was not taken into account in the above calculations. Namely, there is a natural interplay between the power spectrum magnitude and the value of the frequency exponent, $\gamma.$ As $\gamma$ deviates from unity, $\varkappa$ in Eq.~(\ref{sf4}) becomes dimensional, $\varkappa \to \varkappa (f_*)^{\delta},$ where $\delta \equiv \gamma - 1$ and $f_*$ is a frequency parameter. Its value is determined by the physical process responsible for the deviation, and is naturally a large number when measured in hertz. For instance, in the case of charge carrier--phonon interaction, $f_*$ is the ratio of the sound speed to the lattice constant, $f_* \approx 10^{13}$ \cite{kazakov2}. Therefore, already a comparatively small deviation of $\gamma$ can noticeably change the magnitude of $S(f).$ At the same time, $\gamma$ is not easy to measure accurately. Thus, {\it e.g.}, a 0.05 error in $\gamma$ will bring a factor of $10^{13 \cdot 0.05}\approx 5$ in $\varkappa.$

\begin{figure}
\includegraphics[width=0.48\textwidth]{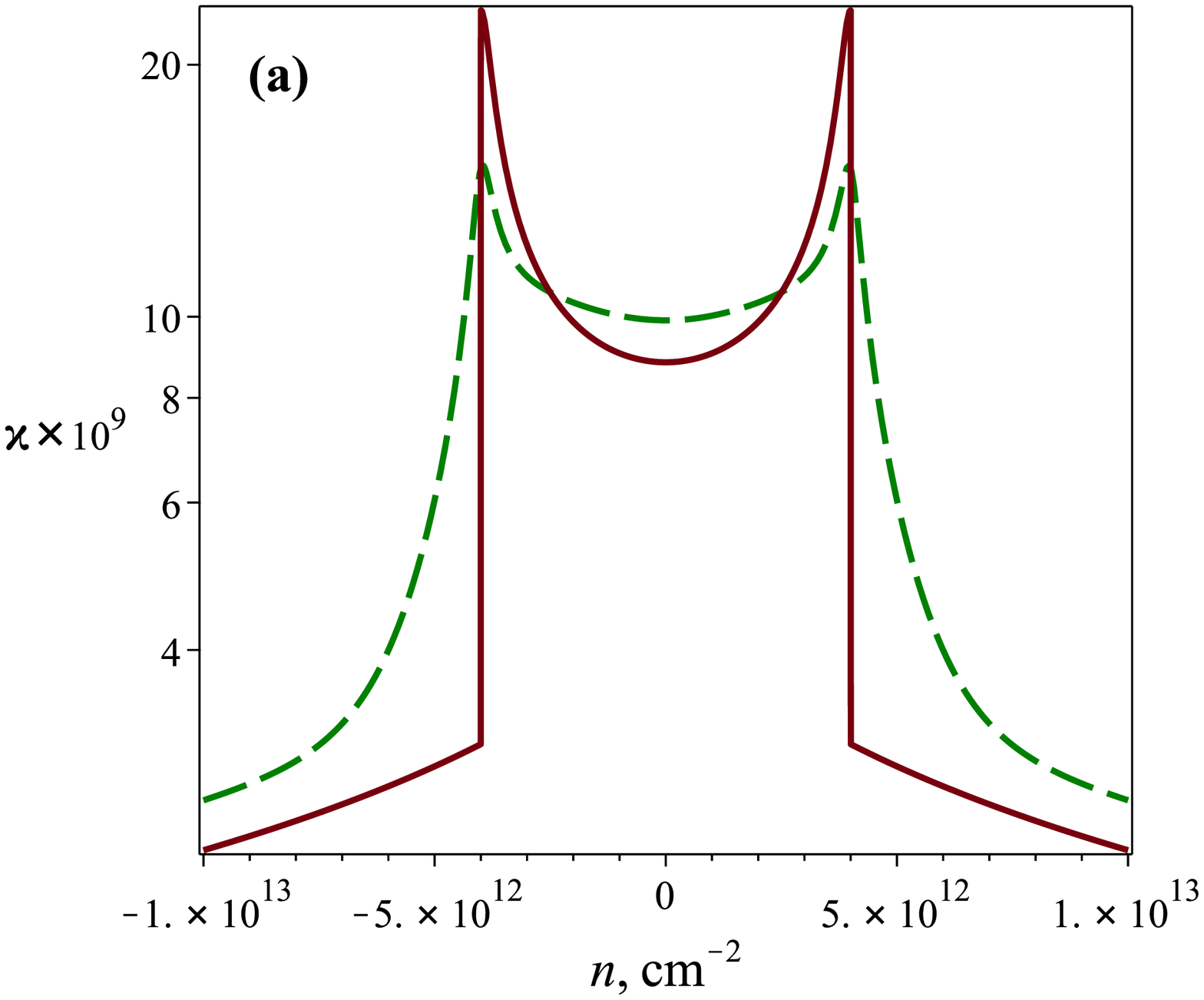}
\includegraphics[width=0.48\textwidth]{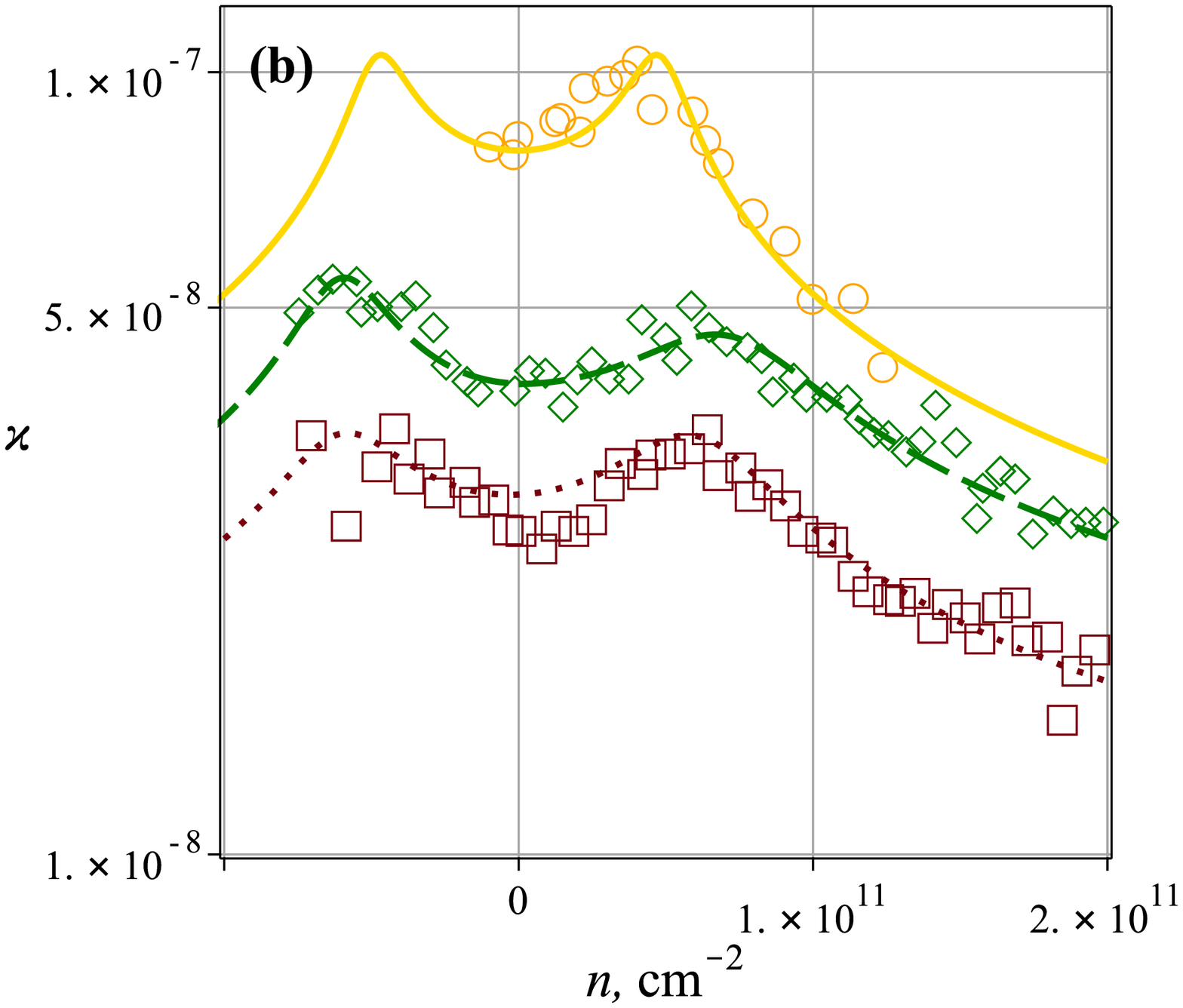}
\caption{(a) M-shaped $\varkappa(n)$s obtained as described in the text without smearing (solid line), and smeared using a model weight function $(1+\exp(n-n_{\rm i})/\delta n)^{-1}$ (dashed line). $n_{\rm i}=4\times 10^{12}\,$cm$^{-2},$ $\delta n= 10^{12}\,$cm$^{-2}.$ (b) Same in a suspended graphene flake of Ref.~\cite{pal2011} at $T=300\,$K (solid line, circles), $T=250\,$K (dashed line, diamonds), and $T=200\,$K (dotted line, boxes). Suspension reduces the effect of substrate disorder, allowing thereby resolution of the M-shape.}\label{fig3}
\end{figure}

As is well-know, the experimentally measured $\varkappa(n)$s are often M-shaped \cite{lin2008,pal2009,heller2010,zhang2011,pal2011,kaverzin2012}. From the present theory standpoint, this readily follows from another widely recognized fact that at sufficiently small charge carrier densities, the charge transport is mixed because both types of charge carriers are present simultaneously. Specifically, as the electron density drops below $n=n_{\rm i} \approx 10^{11}\,$cm$^{-2}-10^{12}\,$cm$^{-2},$ graphene becomes inhomogeneous because of the occurrence of hole ``puddles,'' which grow and coalesce as the electron density decreases, until they fill up the whole sample at $n=-n_{\rm i}$ \cite{hwang2007}. The fact that the average hole (electron) density vanishes not at $n=0,$ but at $n=n_{\rm i}$ ($n=-n_{\rm i}$) means that the left (right) wing of the curve $\varkappa(n)$ is to be shifted to the right (to the left) by the amount of $n_{\rm i},$ and the true $\varkappa$ be obtained by adding the two contributions. Strictly speaking, in view of the sample inhomogeneity, $n\in (-n_{\rm i},n_{\rm i})$ no longer has the meaning of charge carrier density (it is to be considered rather as a measure of the gate voltage), and likewise Eq.~(\ref{sf4}) is not applicable because $R(\bm{r},\bm{Q})$ in Eq.~(\ref{sf1}) does depend on $\bm{r}$ within the sample. But if the puddles are distributed approximately uniformly throughout, and under the assumption that the graphene band structure remains unchanged, Eq.~(\ref{sf4}) is approximately valid. One then obtains curves with M-shaped peaks as shown in Fig.~\ref{fig3}(a). To account for the density fluctuations, $\varkappa(n)$ is to be smeared on the scale $\delta n.$ The resulting $\varkappa(n)$s are superimposed on the experimental data of Ref.~\cite{pal2011} in Fig.~\ref{fig3}(b) where theoretical curves are obtained using the value $n_{\rm i}=5\times 10^{10}\,$cm$^{-2}$ as estimated in Ref.~\cite{pal2011}, and $\delta n$ chosen $10^{11}\,$cm$^{-2}$ to $10^{12}\,$cm$^{-2}$ so as to obtain best fits to the experimental data (the latter is read off from figure 2(c3) of Ref.~\cite{pal2011} where the sets of marks for different temperatures were separated by shifting them vertically; unfortunately, the authors of Ref.~\cite{pal2011} were unable to specify the shifts,\footnote{A.N.~Pal, {\it private communication.}} which does not allow accurate comparison of the theoretical and observed noise magnitudes). It is seen that as in the case of $\Lambda$-shapes in Fig.~\ref{fig2}(b), the theoretical M-shaped curves are congruent with the experimentally observed.

To summarize, account of the detailed band structure of graphene raised the quantum bound \cite{kazakov3} on the $1/f$ noise by a factor of $7/2,$ nearing the experimentally observed values. This result, together with the demonstrated congruence of the calculated and measured noise magnitudes versus charge carrier density, sharpens the question of the flicker noise origin in graphene. It might be readily resolved if estimates of contributions from other possible noise sources were available. However, a common view on the flicker noise as originating from fluctuation of the electric charge distribution around the graphene film still lacks a consistent theory, remaining purely phenomenological. For instance, a variant of this approach suggested in Ref.~\cite{pal2011}, which considers the charge exchange between graphene and its environment as well as the charge rearrangement within the environment, involves four freely adjustable parameters including the noise magnitude. In these circumstances, the issue can be settled only by measurements which would allow sufficiently accurate determination of the frequency exponent to reduce the uncertainty in the noise magnitude, and to verify its dependence on the sample geometry predicted by Eq.~(\ref{sf4}).

\end{document}